\documentclass{article}

\usepackage{PRIMEarxiv}

\usepackage[utf8]{inputenc} %
\usepackage[T1]{fontenc}    %
\usepackage{hyperref}       %
\usepackage{url}            %
\usepackage{booktabs}       %
\usepackage{amsfonts}       %
\usepackage{nicefrac}       %
\usepackage{microtype}      %
\usepackage{lipsum}
\usepackage{fancyhdr}       %
\usepackage{graphicx}       %
\usepackage{subcaption}
\usepackage{caption}       %
\graphicspath{{figures/}}     %
\usepackage[table,xcdraw]{xcolor}
\usepackage{array}
\usepackage{colortbl}
\usepackage{placeins}
\usepackage{soul,color}
\usepackage{amsmath}

\title{Physics guided dual Self-supervised learning for structure-based materials property prediction
\thanks{\textit{\underline{Citation}}: 
\textbf{Fu, Wei, and Hu. Physics Guided Dual SSL for materials property prediction. DOI:000000/11111.}} 
}

\author{
  Nihang Fu\\
 Department of Computer Science and Engineering\\
  University of South Carolina\\
  Columbia, SC 29201 \\
   \And
  Lai Wei\\
 Department of Computer Science and Engineering\\
  University of South Carolina\\
  Columbia, SC 29201 \\
   \And
 Jianjun Hu *\\
 Department of Computer Science and Engineering\\
  University of South Carolina\\
  Columbia, SC 29201 \\
  \texttt{jianjunh@cse.sc.edu} \\
}

\begin{document}
\maketitle

\begin{abstract}
Deep learning (DL) models have now been widely used for high-performance material property prediction for properties such as formation energy and band gap. However, training such DL models usually requires a large amount of labeled data, which is usually not available for most materials properties such as exfoliation energy and elastic properties. Self-supervised learning (SSL) methods have been proposed to address this data scarcity issue by learning inherent representations from unlabeled data in various research fields. Herein, we present DSSL, a physics-guided Dual SSL framework, for graph neural networks (GNNs) based material property prediction. This hybrid framework combines node-masking based predictive SSL with atomic coordinate perturbation based contrastive SSL strategies, allowing it to learn structural embeddings that capture both local and global information of input crystals. Especially, we propose to use predicting the macroproperty (e.g. elasticity) related microproperty such as atomic stiffness as an additional pretext task to achieve physics-guided pretraining process. We pretrain our DSSL model on the Materials Project database with unlabeled data and finetune it with ten extra datasets with different material properties. The experimental results demonstrate that teaching neural networks some physics using the SSL strategy can bring up to 26.89\% performance improvement compared to the baseline GNN models. Our source code is now freely available at \url{https://github.com/usccolumbia/DSSL}. %

\end{abstract}


\section{Introduction}
\label{sec:intro}

Material property prediction is an important and challenging topic in the field of material design and discovery \cite{le2012quantitative,hu2022piezoelectric,varivoda2023materials}. Traditional experimental methods or the first principle density functional theory (DFT) based simulations have excessive cost or significant computational complexity, to determine material properties. In contrast, machine learning (ML) based methods exhibit both time and cost efficiency, making them suitable for high-throughput screening. Among existing machine learning models for material property prediction, graph neural networks (GNNs) based algorithms have emerged as a widely adopted high-performance technique over the past few years \cite{fung2021benchmarking}, with natural representation of materials as graphs, in which atoms serve as nodes and bonds between different atoms as edges. Xie et al. \cite{xie2018crystal} presented the Crystal Graph Convolutional Neural Network (CGCNN) that utilizes atom types and atom connections within crystals to formulate input graphs and then predict material properties. This method provides universal representations of crystalline materials. Chen et al. \cite{chen2019graph} introduced the MatErials Graph Network (MEGNet) for material property prediction for both molecules and crystals given their atoms, bonds, and global state attributes with a special global state. Louis et al. \cite{louis2020graph} developed the global attention-based Graph Convolutional Neural Network (GATGNN) for predicting inorganic material properties based on the graph neural network, which contains multiple graph-attention layers (GAT) and a global attention layer with a speical residual connection to enable deep networks. In addition to these models, there exist many other GNN-based models for material property predictions, such as ALIGNN \cite{choudhary2021atomistic} and DeeperGATGNN \cite{omee2022scalable}. However, as a type of supervised machine learning models, these GNN models face two major challenges when predicting material properties like other supervised ML methods: the scarcity of large-scale labeled data and the suboptimal generalization performance of models, which poses difficulties in the prediction of material properties for new materials \cite{schrier2023pursuit}.%
The conventional deep neural networks (DNNs) for material property prediction are usually trained with supervised learning, where the performance has grown roughly logarithmically with the annotated dataset size \cite{fung2021benchmarking}. The cost of obtaining such data annotations experimentally or computationally has been proven a bottleneck to hinder the scalable progress of these network models. It is a barrier to the development of supervised learning DNNs, especially for the applications that data and annotations are rare, costly, or time-consuming to collect \cite{ericsson2022self}. 

The advent of self-supervised learning (SSL) \cite{liu2021self} liberates deep learning models from the dependencies on large amounts of labeled data. By leveraging input data itself as supervision, SSL has demonstrated superior performance for downstream tasks with its strong representation learning. For example, Cao et al. \cite{cao2023moformer} showed that SSL-based transformer models can improve property prediction performance by pretraining their MOFormer model via maximizing the cross-correlation matrix between its structure-agnostic representations and structure-based CGCNN encoding while also achieving higher data-efficiency compared to structure-only GNN models. Kolluru et al. \cite{kolluru2022transfer} utilized SSL on an attention-based GNN to transfer knowledge across 3D atomic systems from a catalyst dataset to a small molecule database, aiming to achieve generalization across different domains. This approach demonstrated superior performance compared to the non-pretraining method when applied to out-of-domain data.
SSL DL methods typically contain two stages. The first stage, or the pretraining stage, uses pretext tasks to extract valuable insight from the input, which is beneficial for downstream tasks. Subsequently, the second stage, or the finetuning stage, focuses on transferring knowledge acquired in the pretraining stage to effectively address some downstream tasks following conventional supervised training. However, in contrast to traditional supervised learning methods, which rely on a substantial amount of labeled data for supervision, SSL methods provide an alternative that acquires general and valid representations from large amounts of unlabeled data during the pretraining stage for subsequent downstream tasks. During the pretraining stage, SSL proposes various pretext tasks for the network to solve using objective functions similar to those used in supervised learning. However, the labels required for formulating the objective function are automatically obtained from the input data itself (e.g., material structures), obviating the need for human labeling \cite{doersch2017multi}. SSL is usually classified into three main categories: contrastive methods, generative methods, and predictive methods. Using the GNN as an example, the contrastive methods contrast the representations generated from different augmentations of the input graph, and the similarities between augmented sample pairs are used as self-supervision signals. In contrast, the generative SSL methods use the part information of the input graph such as its node or edge attributes and structures, embedded in the graph, as the self-supervision signals. Generative SSL methods are usually based on the pretext tasks like reconstruction. Finally, the predictive SSL methods generally generate labels using some simple statistical analysis from the input graph and design prediction-based tasks based on generated labels to understand the relationship between data and labels \cite{wu2021self}. 

There have been some researches leveraging GNNs with SSL to predict material properties. Magar et al. \cite{magar2022crystal} introduced the Crystal Twins (CT) network based on CGCNN to adapt contrastive learning methods with Barlow Twins \cite{zbontar2021barlow} and Simsiamese loss \cite{chen2021exploring} functions to predict material properties. They used atomic coordinate perturbation and atom and bond masking to get augmented samples for any given input material graph. %
However, despite surpassing the performance of the backbone network (CGCNN), their model fails to outperform state-of-the-art (SOTA) supervised methods due to the limited effectiveness of using simple contrastive SSL. This work maximizes the similarity between two graph embeddings from two perturbed input graphs, which only captures global graph information, but neglects the local information (atoms or bonds). Das et al. \cite{das2023crysgnn} used a hybrid SSL approach by combining the generative SSL via the generic atomic features (including group number, period number, electronegativity, covalent radius, valence electrons, first ionization energy, electron affinity, block, atomic volume) and bond reconstruction and a contrastive SSL
component that maximizes the similarity between samples from the same crystal systems and minimize the similarity of sample pairs from different crystal systems. An additional predictive SSL component is also introduced using the space group prediction as the pretext task. With these three pretext tasks, the goal is to distill important structure information as a learned latent representation for downstream property prediction. Different from conventional SSL methods that share the weights of the network between two stages to carry out downstream tasks, Das' work used distilled information from the first stage as features and combined with existing supervised ML methods to improve their performance based on other compositional or structural features. However, this work only considers structural graph information and a large number of generic atomic properties during the pretraining stage, which does not necessarily improve the model performance in downstream tasks. Another recent work from Suzuki et al. \cite{suzuki2022self} applied contrastive SSL to learn a more powerful material representation by fusing the CGCNN-based graph embedding with the XRD-based embedding for a given material. They found that their material embeddings can capture the structure-functionality relationships among materials and materials concepts. Large-scale SSL has also been proposed recently for materials property prediction by Shoghi et al. \cite{shoghi2023molecules}, which includes energy and force prediction as the pre-training tasks. Their models pretrained on a dataset consisting of about 120 million organic and inorganic systems from OC20 \cite{OC20}, OC22 \cite{OC22}, ANI-1x \cite{ANI-1}, and Transition-1x \cite{Transition1x}, have shown good performance for downstream property predictions. 

In this paper, our objective is to tackle the aforementioned problems and challenges associated with the existing supervised or self-supervised methods for material property predictions by harnessing the power of physics-guided self-supervised learning. Our methods are inspired by the fact that it is the ensemble of the microproperties of local atoms that determines the macroproperties of a material. For example, the elastic property (macroproperty) of materials is strongly dependent on the atomic stiffness \cite{jin2023atomic}, a microproperty, which characterizes the resistance of an atom to volumetric strain. While a generic GNN such as CGCNN can be trained to map a crystal structure into its macroproperty, the learned atomic and bond embeddings do not necessarily reflect or follow the physical rules or constraints, so the trained models have low prediction performance or low generalization capability as the models do not learn sufficient physics during the training process other than minimizing the macroproperty prediction error. Our idea here is to exploit SSL to inject the physics concepts into the neural network training process so that the learned atomic and bond embeddings reflect the physics of macroproperties.

Microproperties of materials manifest at the lattice level and are typically examined through diffraction or spectroscopic methods. Examples of microproperties include the lattice constants, its temperature variation, and the amplitudes of atomic thermal vibrations. Instead, macroproperties such as elastic properties, hardness, and dielectric properties are investigated through measurements of materials in bulk \cite{Sirdeshmukh2006MicroAM}.Mcroproperties and macroproperties of materials exhibit a robust correlation, so that changes at the microscale can profoundly impact the overall behavior of materials. As an illustration, valence electrons, found in the outermost shell of an atom and capable of participating in the formation of chemical bonds, represent one of the key atomic microproperties, which is strongly related to the material band gap, a macroproperty that denotes the energy difference between the valence band and the conduction band. When we aim to design a robust neural network model for band gap prediction, it does not make sense to not involve the valence electron physics into the model or training process considering their strong correlation. This is especially true when only limited training data is available, which makes it insufficient to push the network to learn such valence electron physics in the model. Considering an input material graph including atoms, bonds, or even bond angles, the majority of existing machine learning models predict material properties (macroproperties) directly without an effective way to constrain the network training by weighing related physics. However, mapping an input crystal structure to its material properties reliably is not a straightforward process considering the atomic microproperties and their interactions that lead to high-level macroproperties. To leverage the inherent relationship between microproperties and macroproperties, we propose to utilize the SSL framework to push the neural network model to learn the related microproperty physics so that higher property prediction performance can be achieved. This is done by pretraining the neural networks to learn materials embedding using microproperty prediction pretext tasks along with contrastive SSL. This will allow our network models to establish a physic-guided connection between input structures/graphs with their microproperties and the global macroproperties of the material.

In this work, we present DSSL, a two-stage physics-guided GNN-based dual SSL framework for material property prediction. During the pretraining stage, the model is trained using a predictive SSL strategy in combination with a contrastive SSL strategy, making effective use of extensive amounts of unlabeled data selected from the Materials Project (MP) database \cite{10.1063/1.4812323}. During the finetuning stage, we employ ten datasets from the MatBench \cite{dunn2020benchmarking} to finetune the pretrained model obtained from the first stage, enabling the prediction of various material properties as shown in Table \ref{tab:datasets}. Especially, we introduce the physics-guided pretext tasks for pretraining: for elasticity prediction, the atomic stiffness prediction is used as one of its pretext tasks; for band gap prediction, the valence electron prediction is used. 
Additionally, ablation studies are conducted to investigate the effect of different SSL strategies and the effect of the physics-guided pretext task on the downstream macroproperty prediction performance. 

\begin{figure}%
    \centering
    \includegraphics[width=\linewidth]{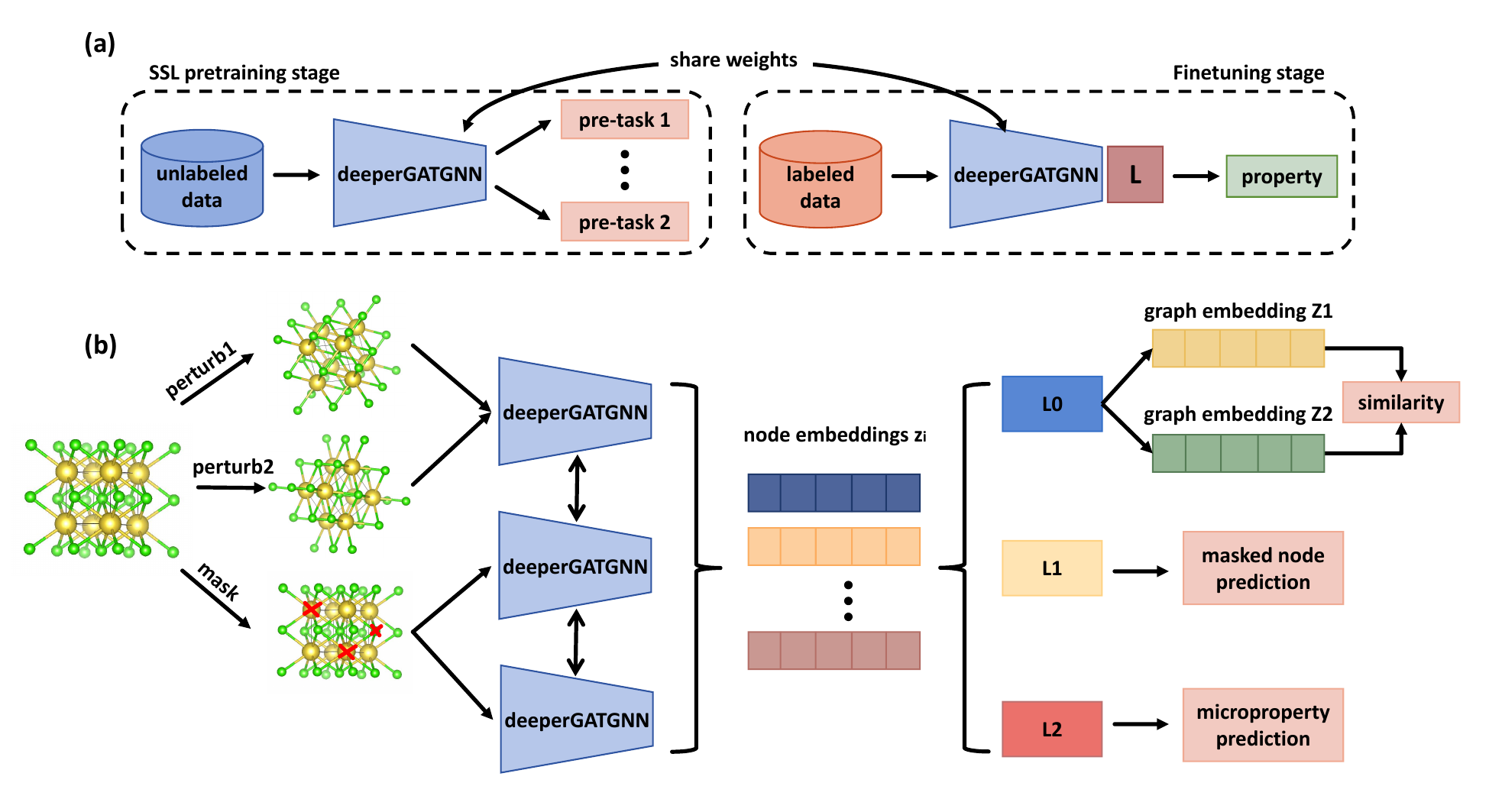}%
    \caption{DSSL Framework. (a) The two-stage DSSL. Within this framework, we employ DeeperGATGNN as the encoder. The pretraining stage trains DeeperGATGNN on an unlabeled database using SSL methods, and the subsequent finetuning follows the training of the original DeeperGATGNN on some labeled datasets but DeeperGATGNN with the shared weights from the pretraining stage. (b) The details of the pretraining stage. When an unlabeled material is input to the network, we apply perturbation and masking techniques to generate augmented data (i.e., two perturbed data and one masked data). Subsequently, these augmented input data are individually fed into DeeperGATGNN for different pretraining tasks.}%
    \label{fig:architecture}%
\end{figure}

\section{Method}
\label{sec:headings}

\subsection{Physics guided SSL Framework for GNN-based material property prediction}
We propose a physics-guided dual SSL GNN-based framework for material property prediction composed of two stages (as shown in Figure \ref{fig:architecture}.(a)). During the pretraining stage, we utilize DeeperGATGNN \cite{omee2022scalable} as the backbone and incorporate three types of SSLs including mask-based generative SSL, contrastive learning SSL, and physics-guided predictive SSL to obtain a pretrained model. In this stage, no extra labeled data is needed, and all pretext tasks force the network to learn features that are useful for the downstream tasks (i.e., material property prediction) in a self-supervised manner. During the finetuning stage, the model is further trained on datasets with labeled properties in a supervised learning way, yielding superior performance compared to models trained from scratch. The weights of the backbone are shared between the pretraining and finetuning stages.

\paragraph{DeeperGATGNN for property prediction}
The state-of-the-art deep learning methods for predicting material properties are based on GNNs given the graphs from material structures. For a material graph $G = \{V, E\}$ where $V$ represents the atom set and $E$ represents the connection relationships between atoms, a basic GNN like DeeperGATGNN updates the node information by aggregating messages from the neighborhood of this node. The message-passing update function of basic GNNs can be expressed as follows: %

\begin{equation}
    \boldsymbol{h}_u^{(k)} = {f_u}^{(k-1)}\biggl(\boldsymbol{h}_u^{(k-1)}, {f_a}^{(k-1)}\bigl(\{h_v^{(k-1)}, \forall v \in \boldsymbol{N}(u)\}\bigr)\biggr)
\end{equation}

where $f_u$ and $f_a$ represent the update function and the aggregate function (i.e., neural networks) and $\boldsymbol{N}(u)$ represents the neighborhood sets of the node $u$. At each iteration $k$, the aggregate function $f_a$ takes the set of embeddings of the nodes in node $u$'s neighborhood $\boldsymbol{N}(u)$ as input and generates a message based on this aggregated neighborhood information. The update function $f_u$ combines this aggregated information with embedding of node $u$ at iteration $k-1$ (i.e., $\boldsymbol{h}_u^{(k-1)}$) to generate the updated embedding $\boldsymbol{h}_u^{k}$. 

In this paper, we use DeeperGATGNN as the backbone to encode the input graphs. Based on the basic GNNs, the DeeperGATGNN introduced a novel global-attention GNN layer with a differentiable group normalization (DGN) and residual skip connections to achieve scalable training of highly deep graph neural network models for materials property prediction \cite{omee2022scalable}.

\paragraph{Mask-based generative SSL} %

Atom masking is one of the SSL strategies we use to capture the local microproperty information (i.e., node information) in the pretraining stage (as shown in the second row of Figure \ref{fig:architecture}.(b)). In the original DeeperGATGNN, the node feature is represented with one-hot encoding regarding different element types. The masking process is to set the node feature of a masked atom as a special vector. In our work, we randomly mask atoms at a 0.15 ratio on a given graph. Then we input the masked graph into DeeperGATGNN to generate node embeddings $z_i$ for all atoms, which are then used to predict the element types of the masked atoms. This is done by the linear layer $L_1$ that follows DeeperGATGNN. So the objective of mask-based prediction is to push the neural network to learn to encode the local atomic context corresponding to each specific element type by the pretext task of predicting the masked atom types. Here the cross-entropy loss is used as the objective function as shown below:

\begin{equation}
L_{mask} = CrossEntropyLoss(L_1(z_i), y) \\
         = -\Sigma_{c=1}^C y_c\ log\ L_1(z_i)_c
\end{equation}

where $z_i$ is the node embedding; $y$ is the target; $C$ is the number of the atom types.

\paragraph{Contrastive SSL} Contrastive learning is one of self-supervised learning approaches to learning to extract meaningful representations by contrasting a pair of instances. Intuitively, contrastive learning can be considered as learning by comparing \cite{le2020contrastive}, which pushes the network to implicitly learn to extract the essential features or representations of the data that do not vary with the instance transformations or augmentations. In this work, we utilize contrastive learning to capture global information (i.e., graph information) during the pretraining stage as global structure perturbations are used to generate the graph embeddings.  As shown in the first row of Figure \ref{fig:architecture}.(b), given an input crystal structure graph, random perturbations are applied to the coordinates of the atoms within the unit cell to generate two augmented structures. Each structure instance is then fed into DeeperGATGNN to generate its graph embedding. After we calculate the node representations $z_i$ for all atoms, we aggregate them using a linear layer $L_0$ to generate a vector representing the entire graph for each of the sample pair, called the graph embeddings $Z_1$ and $Z_2$. The core function of contrastive learning lies in maximizing the similarity between two embeddings $Z_1$ and $Z_2$ of the perturbations of the given material. This method aids the network in acquiring the ability to filter out the unimportant, variable, and noisy information within the material representation, enabling more concentrated emphasis on the crucial and consistent information inherent to the representations of the same material. The objective of our contrastive learning module is to maximize the cosine similarity between these two embeddings of the sample pair, and the loss function is described as:

\begin{equation}
L_{cl} = CosineEmbeddingLoss(Z_1, Z_2, y) = 1-cos(Z_1, Z_2)
\end{equation}

\paragraph{Physics-guided predictive SSL} 
It is well known that microproperties, or the properties of the individual atoms and bonds within a material and their combinations significantly influence the corresponding macroproperties of the material. With this prior knowledge, we propose to use specific microproperties prediction as one of the pretext tasks for SSL during the pretraining stage (as shown in the third row of Figure \ref{fig:architecture}.(b)). This approach aims to enhance the overall macroproperties prediction performance by leveraging the insights that the network learned from predicting the microproperties. In this paper, we consider two specific microproperties, namely, Valence Electrons (VE) and Atomic Stiffness (AS). Valence electrons typically influence macroproperties such as the band gap, while atomic stiffness can impact the elasticity properties of materials. Accordingly, we prepare the Band Gap dataset \cite{jain2013commentary} to evaluate the network performance with valence electrons as one of the pretext tasks and prepare KVRH \cite{dunn2020benchmarking, deJong2015} and GVRH \cite{dunn2020benchmarking, deJong2015} datasets to evaluate the network with atomic stiffness as one of the pretext tasks. The details of the datasets can be found in Section \ref{sec:res}. As VE prediction is a classification task with 17 categories, we employ the cross-entropy loss for its prediction. On the other hand, AS prediction is a regression task, so we choose the L1 loss.

\begin{equation}
L_{micro} =  \left\{ \begin{array}{lcl}
CrossEntropyLoss(L_2(z_i), y), & \mbox{if}
& {VE} \\ |y-L_2(z_i)|, & \mbox{if} & {AS}
\end{array}\right.
\end{equation}

where $z_i$ represents the node embedding of the $i^{th}$ node; $y$ is the ground truth (true VE or true AS).

By combining mask-based generative SSL, contrastive SSL, and predictive SSL with physics-guided pretext tasks, the overall loss for the framework can be written as:
\begin{equation}
L_{pretrain} = \alpha L_{mask} + \beta L_{cl} + \gamma L_{micro}
\end{equation}

\subsection{Evaluation criteria}
We utilize the Mean Absolute Error (MAE) to evaluate the performance of our work, which measures the error between the predicted properties and the corresponding true targets.

\begin{equation}
E = \frac{\Sigma_{n=1}^N |y_n - f(x_n)|}{N}
\end{equation}

where $N$ is the sample size; $x_n$ is the $n^{th}$ material; $f(.)$ represents the network used to predict material properties; $y_n$ and $f(x_n)$ denote the $n^{th}$ target and prediction, respectively.

For a fair comparison with baseline SSL methods, we conduct experiments using ten repetitions of training on the same split training set of HOIP, Lanthanides, Fermi Energy, Formation Energy, and Band Gap datasets and five-fold cross-validation method for JDFT2D, Phonons, Dielectric, GVRH, and KVRH datasets. This allows us to compute the MAE $E_i$ for each run $i$ and subsequently determine the mean $\overline{E}$ and standard deviation $\sigma$ across these values.

\begin{equation} 
 \overline{E} = \frac{1}{K}\Sigma_{i=1}^K\frac{\Sigma_{n=1}^N |y_n - f(x_n)|}{N} 
\end{equation}

\begin{equation} 
\sigma = \sqrt{\frac{1}{K-1} \sum_{i=1}^K (E_i - \overline{E})^2}
\end{equation}

where $K$ is the number of runs (five or ten), $E_i$ represents the MAE score of the $i^{th}$ run.

\section{Results}
\label{sec:res}

\subsection{Datasets}
We initially pretrain our model using the Materials Project (MP) database \cite{materialsproject}, followed by finetuning it on ten distinct datasets whose details are summarized in Table \ref{tab:datasets}.  We have one pretraining dataset, MP, and divide it into the training set and the validation set with a 9:1 ratio. We have a total of ten finetuning datasets. Here, five of them are used in previously published work, like Crystal Twin\cite{magar2022crystal}, CGCNN \cite{xie2018crystal}, and OGCNN \cite{karamad2020orbital}, while the remaining five datasets are sourced from the Matbench suites \cite{dunn2020benchmarking}. For the first five datasets (HOIP, Lanthanides, Fermi Energy, Formation Energy, and Band Gap datasets), we split them into training, validation, and test sets with a 6:2:2 ratio, and then repeat the training process ten times to evaluate the network performance. Furthermore, we employ a five-fold cross-validation method on the remaining five datasets (JDFT2D, Phonons, Dielectric, GVRH, and KVRH datasets). In both cases, we present the performance through the $\overline{E}$ and $\sigma$ calculations.

\begin{table}[]
\centering
\setlength{\abovecaptionskip}{0.2cm}
\caption{Datasets used for both pretraining and finetuning stages.}
\label{tab:datasets}
\begin{tabular}{
>{\columncolor[HTML]{FFFFFF}}c |
>{\columncolor[HTML]{FFFFFF}}c 
>{\columncolor[HTML]{FFFFFF}}c }
\toprule[1.5pt]
\textbf{Dataset} & \textbf{Property}  & \textbf{Sample NO.} \\ \hline \\[-1em]
MP \cite{10.1063/1.4812323}            & -                  & 138,614             \\ \\[-1em]

HOIP \cite{kim2017hybrid}             & Band Gap           & 1,345                \\  \\[-1em]
Lanthanides \cite{pham2017machine}     & Formation Energ    & 4,166                \\  \\[-1em]
Fermi Energy \cite{jain2013commentary}   & Fermi energy       & 26,447               \\  \\[-1em]
Formation Energy \cite{jain2013commentary}  & Formation Energ    & 26,741               \\  \\[-1em]
Band Gap \cite{jain2013commentary}      & Band Gap           & 27,110               \\  \\[-1em]
JDFT2D \cite{dunn2020benchmarking}          & Exfoliation Energy & 636                 \\  \\[-1em]
Phonons \cite{dunn2020benchmarking, petretto2018high}         & Last Phdos Peak     & 1,265                \\  \\[-1em]
Dielectric \cite{dunn2020benchmarking, petousis2017high}      & Refractive Index    & 4,764                \\  \\[-1em]
GVRH \cite{dunn2020benchmarking, deJong2015}            & Shear Modulus      & 10,987               \\  \\[-1em]
KVRH \cite{dunn2020benchmarking, deJong2015}            & Bulk Modulus       & 10,987               \\  
\bottomrule[1.5pt]
\end{tabular}
\end{table}

\subsection{Experimental Analysis}

In this section, we conducted a comprehensive evaluation of our models on different datasets and compared them with existing models. Additionally, we designed a set of ablation studies to assess the effectiveness of various SSL strategies implemented in our model to analyze these individual components and unravel their distinct contributions to the overall performance.

\begin{table}[th]
\centering
\caption{Performance comparison on MatBench datasets: JDFT2D (meV/atom), Phonons (1/cm), Dielectric (no unit), GVRH (GPa), and KVRH (GPa).}
\label{tab:comp_matbench}
\begin{tabular}{c|ccccccc}
\toprule[1.5pt]
   & JDFT2D                 & Phonons                & Dielectric             & GVRH                   & KVRH                   \\ \hline
\multicolumn{1}{c|}{CGCNN}                 & 49.24 ± 11.58          & 57.36 ± 12.31          & 0.599 ± 0.083          & 0.089 ± 0.001          & 0.071 ± 0.002          \\
\multicolumn{1}{c|}{CT\_Barlow}            & 46.79 ± 19.92          & 50.33 ± 08.88          & 0.434 ± 0.100          & 0.086 ± 0.004            & 0.067 ± 0.003            \\
\multicolumn{1}{c|}{CT\_SimSiam}           & 48.38 ± 18.68          & 48.86 ± 07.69          & 0.417 ± 0.079          & 0.087 ± 0.003            & 0.067 ± 0.003            \\
\multicolumn{1}{c|}{ALIGNN}                & 43.42 ± 08.95          & 29.53 ± 02.11          & 0.345 ± 0.087          & \textbf{0.071 ± 0.001} & \textbf{0.057 ± 0.003} \\
\multicolumn{1}{c|}{DeeperGATGNN}     & 48.11 ± 08.86          & 39.08 ± 04.69          & 0.393 ± 0.070          & 0.095 ± 0.003          & 0.079 ± 0.001          \\ \hline
\multicolumn{1}{c|}{DSSL} & \textbf{36.29 ± 05.26} & \textbf{28.57 ± 03.09} & \textbf{0.342 ± 0.041} & 0.084 ± 0.003        & 0.068 ± 0.002          \\   
\multicolumn{1}{c|}{DSSL+AS} & - & - & - & 0.083 ± 0.003         & 0.066 ± 0.002          \\   
\bottomrule[1.5pt]                                    
\end{tabular}
\end{table}

\paragraph{Performance comparison}
We first compare the mean $\overline{E}$ and standard deviation $\sigma$ of the MAEs of different models (see Table \ref{tab:comp_matbench} and Table \ref{tab:comp_5datasets}). In Table \ref{tab:comp_matbench}, we compare the results of our models over five datasets from the Matbench for predicting the following material properties: exfoliation energy (JDFT2D dataset), frequency of the highest frequency optical phonon mode peak (Phonons dataset), refractive index (Dielectric dataset), and elastic properties (GVRH and KVRH datasets). In this table, the results of other networks are from MatBench, and the best results are shown in boldface. For our DSSL, the mean and the standard deviation are calculated over five folds following the MatBench. For two datasets with a smaller number of samples (JDFT2D and Phonons), previous models, i.e., CGCNN, CT, and ALIGNN, have an average prediction performance exceeding 40 meV/atom and 30/cm with a considerable corresponding $\sigma$. It indicates the instability of the model performances on these datasets. The MAEs of the DSSL model are 36.29 meV/atom and 28.57/cm, respectively, which outperforms previously published models and achieves a maximum improvement of 26.30\% and a minimum improvement of 16.42\% on the JDFT2D dataset. For the Phonons dataset, it demonstrates a maximum improvement of 50.20\% compared to CGCNN, and a minimum improvement of 3.25\% compared to ALIGNN. Then, for the Dielectric dataset, DSSL has the MAE of 0.342 and also surpasses previous models with an average improvement of 19.19\%. Although DSSL may not surpass the performance of ALIGNN on two large datasets (GVRH and KVRH), it is crucial to acknowledge that this outcome could be influenced by the suboptimal performance of the basic network (DeeperGATGNN) employed. Notably, the original DeeperGATGNN performs less effectively than other models on the GVRH and KVRH datasets. However, despite this limitation, DSSL succeeds in enhancing the original DeeperGATGNN model, surpassing the performance of CGCNN and CT models. In particular, as results of DSSL row shows, DSSL consistently enhances the performance of the original DeeperGATGNN with improvements of 24.57\%, 26.89\%, 12.98\%, 11.58\%, and 13.92\% on 5 datasets, respectively. We also indicate the results of DSSL with the prediction of atomic stiffness as one of the pretext tasks during the pretraining stage (marked as DSSL+AS). Given the strong correlation between atomic stiffness and elastic properties, AS effectively guides the prediction of the shear modulus and bulk modulus, achieving average MAEs of 0.083 GPa and 0.066 GPa, improvements of 12.63\% and 16.46\% compared to the original DeeperGATGNN, and improvements of 1.19\% and 2.94\% compared to DSSL on the GVRH and KVRH datasets, respectively.

\begin{figure}%
    \centering
    \subfloat[\centering MP vs Formation Energy]{{\includegraphics[width=0.45\linewidth]{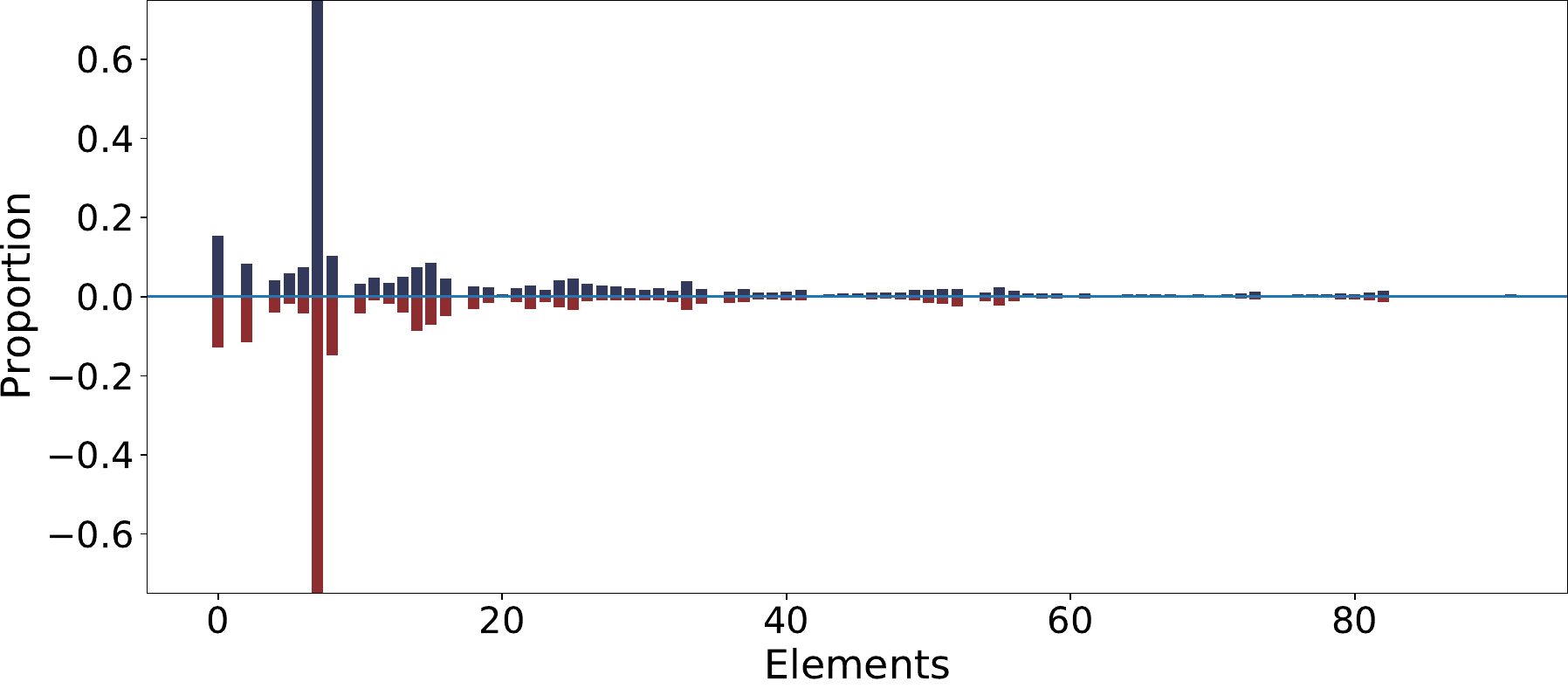} }}%
    \qquad
    \subfloat[\centering MP vs Band Gap]{{\includegraphics[width=0.45\linewidth]{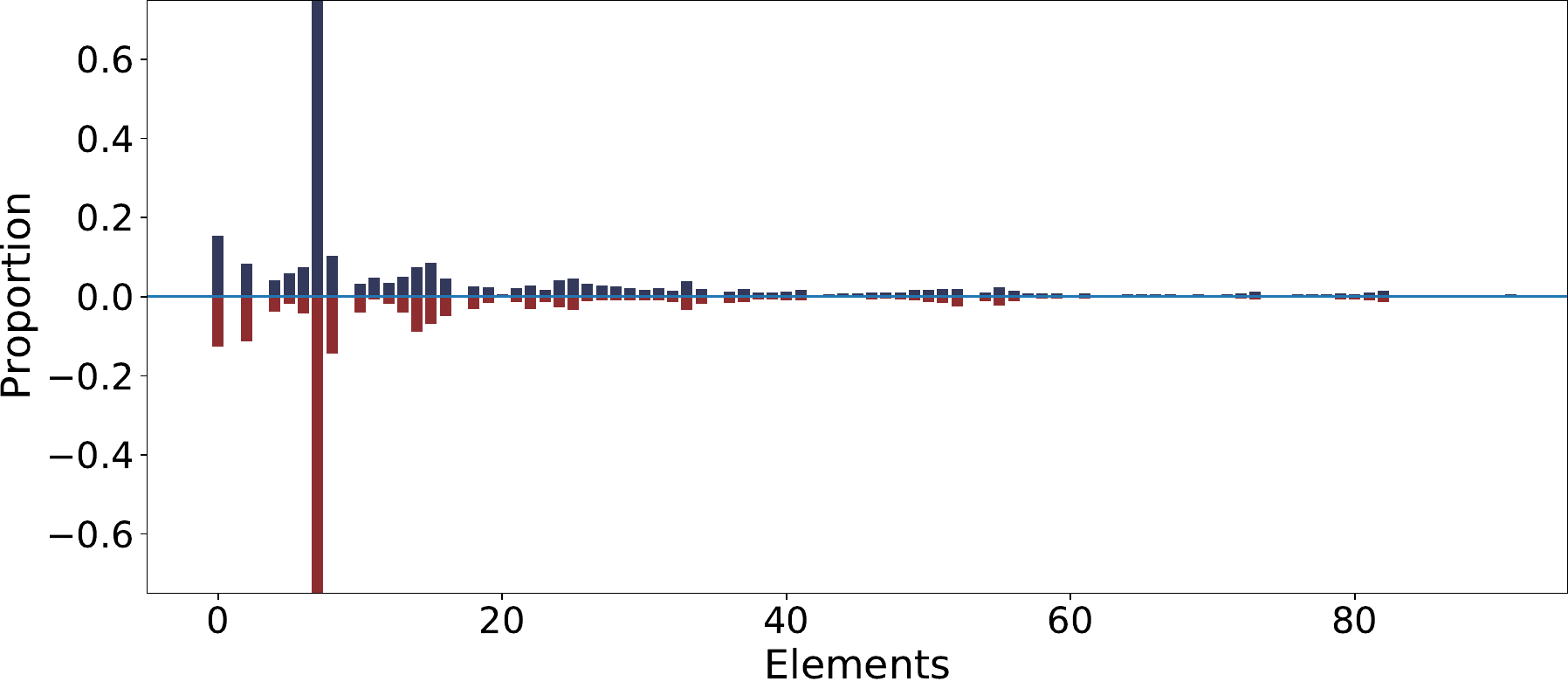} }}%
    \qquad
    \subfloat[\centering MP vs HOIP]{{\includegraphics[width=0.45\linewidth]{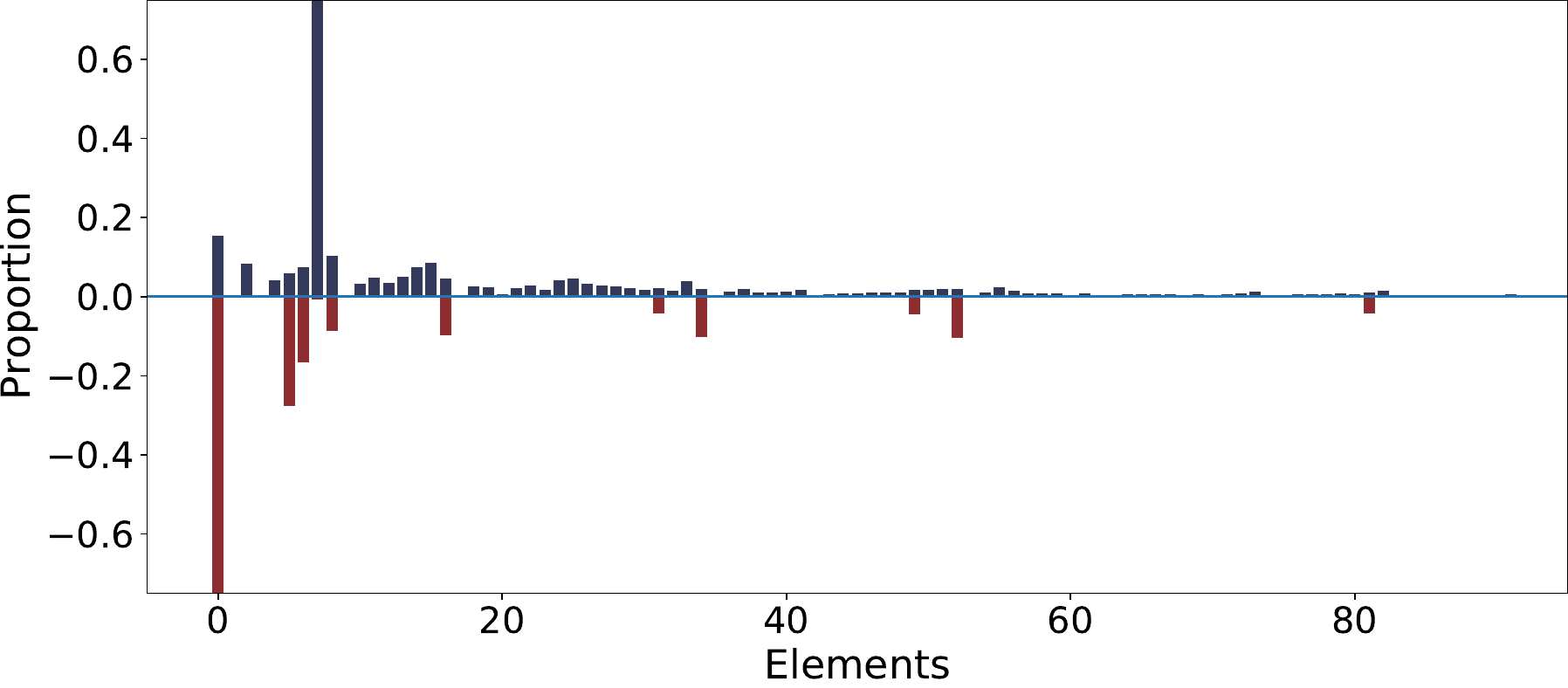} }}%
    \qquad
    \subfloat[\centering MP vs Lanthanides]{{\includegraphics[width=0.45\linewidth]{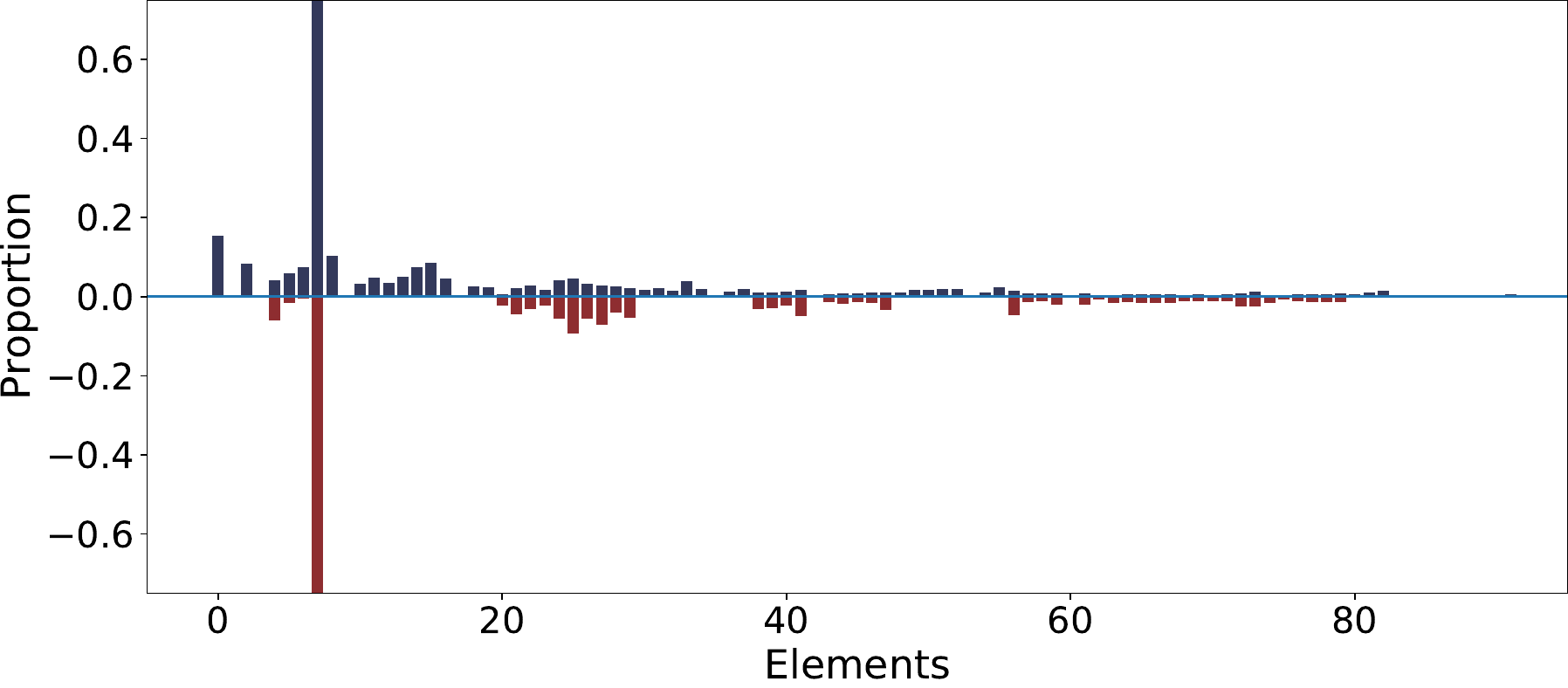} }}%
    \caption{Comparison of element distributions in different datasets. The x-axis represents different element types (e.g., H is 1), and the y-axis represents the proportion of each element in the dataset. (a) MP vs Formation Energy. The blue one represents the MP dataset, while the red one is the Formation Energy dataset. (b) MP vs Band Gap. The blue one represents the MP dataset, while the red one is the Band Gap dataset. (c) MP vs HOIP. The blue one represents the MP dataset, while the red one is the HOIP dataset. (a) MP vs Lanthanides. The blue one represents the MP dataset, while the red one is the Lanthanides dataset.}%
    \label{fig:distribution}%
\end{figure}

\begin{table}[bh]
\centering
\caption{Performance comparison on extra five datasets: Fermi Energy (eV), Formation Energy (eV/atom), Band Gap (eV), HOIP (eV), and Lanthanides (eV/atom).}
\label{tab:comp_5datasets}
\begin{tabular}{c|ccccc}
\toprule[1.5pt]
 & Fermi Energy           & Formation Energy                 & Band Gap                   & HOIP        & Lanthanides   \\ \hline
GIN \cite{XIAO2024112619}                           & 0.605 ± 0.015          & 0.109 ± 0.007          & 0.601 ± 0.038          & 0.666 ± 0.123          & 0.197 ± 0.038          \\
CGCNN                          & 0.400 ± 0.003          & 0.040 ± 0.001          & 0.369 ± 0.003          & 0.170 ± 0.013          & 0.080 ± 0.003          \\
OGCNN \cite{karamad2020orbital}                         & 0.446 ± 0.018          & 0.035 ± 0.001          & 0.353 ± 0.008          & 0.164 ± 0.013          & 0.072 ± 0.002          \\
CIN\_Barlow                    & 0.478 ± 0.125          & 0.085 ± 0.003          & 0.337 ± 0.004          & 0.395 ± 0.007          & 0.094 ± 0.000          \\
CT\_Barlow                     & 0.399 ± 0.004          & 0.025 ± 0.001          & 0.328 ± 0.002          & 0.153 ± 0.003          & 0.058 ± 0.001          \\
CT\_SimSiam                    & 0.384 ± 0.004          & 0.024 ± 0.001          & 0.302 ± 0.001          & \textbf{0.140 ± 0.004} & \textbf{0.054 ± 0.001} \\
DeeperGATGNN             & 0.374 ± 0.014          & 0.023 ± 0.001          & 0.300 ± 0.011          & 0.214 ± 0.017          & 0.105 ± 0.008          \\\hline
DSSL          & \textbf{0.351 ± 0.005} &  \textbf{0.021 ± 0.001}          & \textbf{0.289 ± 0.005} & 0.232 ± 0.015          & 0.154 ± 0.015          \\
\multicolumn{1}{c|}{DSSL+VE} & - &  0.022 ± 0.001         & \textbf{0.283 ± 0.005}&-&-          \\   
\bottomrule[1.5pt]
\end{tabular}
\end{table}
In Table \ref{tab:comp_5datasets}, we compare the results of our DSSL model with those previously published models that predict material properties, specifically focusing on properties including formation energy, fermi energy, and band gap. The results of other networks in this table are from Crystal Twin \cite{magar2022crystal} and the best results are shown in boldface. On the Fermi Energy, Formation Energy, and Band Gap datasets, DSSL has the MAEs of 0.351 eV, 0.021 eV/atom, and 0.289 eV on these three datasets, respectively, and demonstrates improvements compared to the original DeeperGATGNN with improvements of 6.15\%, 8.70\%, and 3.67\%, respectively. Except for the original DeeperGATGNN, DSSL also outperforms other networks on the Fermi Energy, Formation Energy, and Band Gap datasets, which can achieve a maximum improvement of 41.98\%, 80.73\%, and 51.91\%, and a minimum improvement of 8.59\%, 12.50\%, and 4.30\% on these three datasets, respectively. However, the situation differs on the HOIP and Lanthanides datasets. DSSL not only performs less favorably than the previous networks but also fails to surpass the original DeeperGATGNN on these two datasets. To investigate the cause behind this abnormal phenomenon, we plot the bar figures to compare the element distributions between the pretraining dataset and different finetuning datasets as shown in Figure \ref{fig:distribution}. We count the number of each element in different datasets and normalize them to plot this figure. The primary purpose of pretraining is to enhance the network's understanding of the latent content of the input materials. However, in cases where there is a substantial difference between the distributions of the finetuning data and the pretraining data, the pretraining process fails to enhance the inherent understanding of materials for the finetuning stage; instead, it hinders this understanding. As depicted in Figure \ref{fig:distribution}.(a) and (b), the data distributions between the MP dataset and the Formation Energy dataset or between the MP dataset and the Band Gap dataset exhibit a great similarity. Consequently, the pretraining stage proves beneficial for the subsequent finetuning stage, and our DSSL not only outperforms other SOTA networks on these datasets but also improves the performance of the original DeeperGATGNN. However, Figure \ref{fig:distribution}.(c) and (d) reveal significantly disparate data distributions between the pretraining data and the finetuning data. The substantial difference in data distribution presents challenges in realizing enhanced results during the finetuning stage based on the pretraining stage resulting in poor predictions. We also present the results of DSSL with the prediction of valence electron as one of the pretext tasks during the pretraining stage (marked as DSSL+VE). Given the strong correlation between valence electron and band gap, VE effectively guides the prediction of the band gap, achieving an average MAE of 0.283 eV on the Band Gap dataset and an improvement of 5.67\% compared to the original DeeperGATGNN and 2.08\% compared to DSSL. However, since the formation energy lacks a strong relationship with valence electron, there is no improvement observed using DSSL+VE on this dataset.

\begin{figure}%
    \centering
    \subfloat[\centering Atomic Stiffness]{{\includegraphics[width=0.45\linewidth]{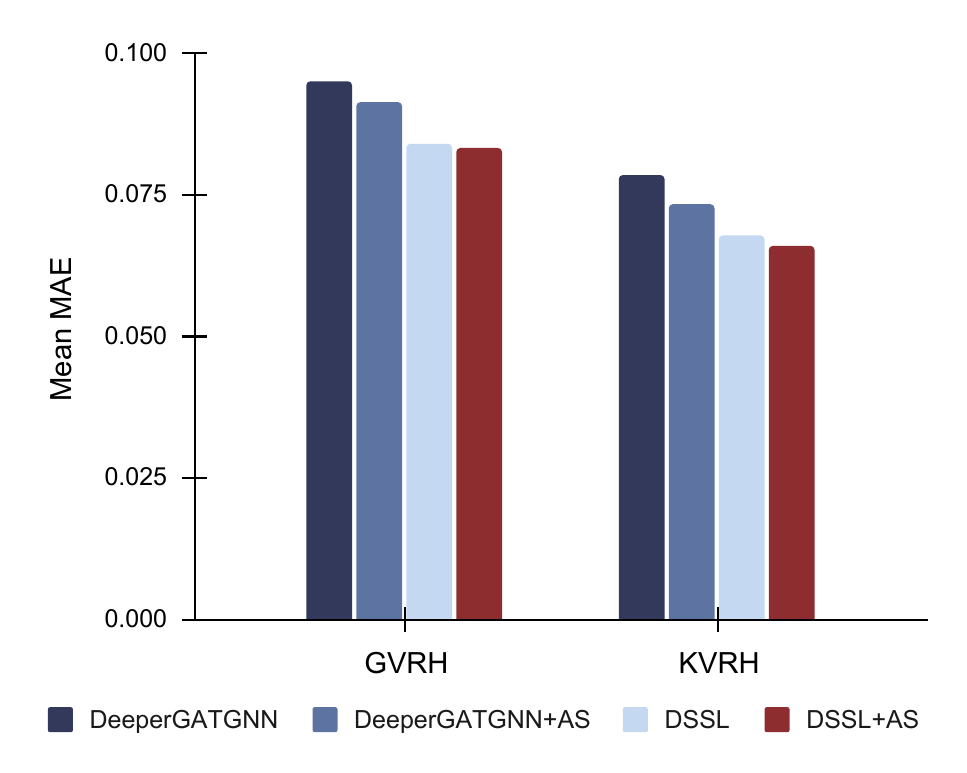} }}%
    \qquad
    \subfloat[\centering Valance Electron]{{\includegraphics[width=0.45\linewidth]{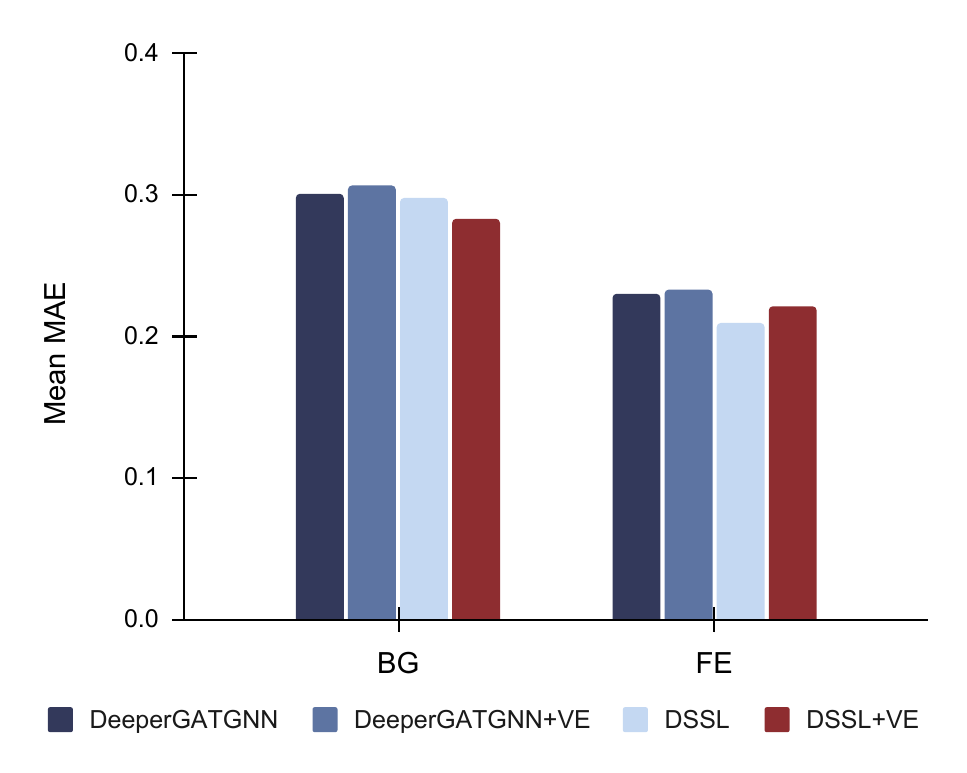} }}%
    \caption{Performance comparison on the incorporation of various microproperties into networks. (a) The model performance on GVRH and KVRH datasets using original DeeperGATGNN and DSSL and using atomic stiffness as a part of the input for DeeperGATGNN and a pre-task for DSSL, respectively. (b) The model performance on Band Gap (BG) and Formation Energy (FE) datasets using original deeperGARGNN and DSSL and using valence electron as a part of the input for DeeperGATGNN and DSSL and a pre-task for DSSL, respectively. For the Formation Energy dataset, we expanded the Mean MAE ten times so that it could be clear in the figure. }%
    \label{fig:micro}%
\end{figure}

We then design the experiments to compare the model performances with and without the help of microproperties based pretext tasks in our DSSL framework as shown in Figure \ref{fig:micro}. We include two distinct microproperties as examples in this experiment: Valance Electron (VE), as introduced in Section \ref{sec:intro}, which is associated with the band gap property, and Atomic Stiffness (AS). %
For AS, we employ two datasets (GVRH and KVRH datasets) with properties related to elasticity (shear modulus and bulk modulus) to examine the model prediction performance. Additionally, for VE, we use the Band Gap dataset and the Formation Energy dataset, where the band gap is one of the macroproperties that is strongly correlated with valence electrons while the formation energy does not have a relationship with VE as strongly as the former one. So here, we use the Band Gap dataset to check the model's band gap prediction performance when the valence electron is introduced to the models and use the Formation Energy dataset as a counterexample to investigate the performance impact when the introduced microproperty lacks a strong relationship with the macroproperty (formation energy) that we want to predict. For each microproperty, we compare the original DeeperGATGNN, the approach of directly adding microproperties to the node attributes of the input graph (marked as "DeeperGATGNN+microproperty" in Figure \ref{fig:micro}), DSSL, and DSSL with the prediction of microproperties as a pretext task during the pretraining stage (marked as "DSSL+microproperty" in Figure \ref{fig:micro}). As illustrated in Figure \ref{fig:micro}.(a), directly adding atomic stiffness to the atomic attributes in the original DeeperGATGNN (DeeperGATGNN+AS) with a mean MAE of 0.091 GPa yields a performance improvement on the original DeeperGATGNN (a mean MAE of 0.095 GPa), but it fails to outperform DSSL (a mean MAE of 0.084 GPa). When we add the atomic stiffness prediction as an additional pretext task, the DSSL+AS achieves a mean MAE of 0.083 GPa, which is the best performance compared to three existing baseline methods on both KVRH and GVRH datasets. Then in Figure \ref{fig:micro}.(b), directly adding VE into the atomic attribute of the input graph in the DeeperGATGNN (DeeperGATGNN+VE) has a mean MAE of 0.307 eV, which does not improve the original DeeperGATGNN. However, adding VE as an additional pretext task (DSSL+VE) has a mean MAE of 0.283 eV and improves the performance of DSSL (a mean MAE of 0.289 eV) on the Band Gap dataset. Conversely, on the Formation Energy dataset, VE fails to improve both DeeperGATGNN and DSSL, leading to a deterioration in their performance. This phenomenon indicates the importance of selecting compatible microproperties to guide the prediction of macroproperties and emphasizes the need for a purposeful physics-guided choice instead of a random set of microproperties in this process as done previously \cite{das2023crysgnn}.

\paragraph{Ablation Study}
To investigate the complementary relationship among different SSL methods, we conduct a comprehensive ablation study for our DSSL algorithm. 
As shown in Table \ref{tab:comp_ssl}, we trained the original DeeperGATGNN (Original), DeeperGATGNN with only mask-based SSL (Mask), DeeperGATGNN with only contrastive SSL (CL), and our DSSL on six different datasets. On the GVRH, Band Gap, Formation Energy, and Fermi Energy datasets, mask-based SSL strategies have performance improvements of different degrees compared to the Original models.  While the contrastive SSL model on its own does not yield satisfactory results, when combined with mask-based SSL (i.e., DSSL), they synergize to establish a complementary relationship, resulting in favorable outcomes. DSSL can achieve the best performance with improvements on the original DeeperGATGNN of 11.58\%, 3.67\%, and 8.70\%, respectively. Moreover, DSSL+microproperties on the corresponding datasets enhance the performance of DSSL, resulting in mean MAEs of 0.083 GPa and 0.283 eV on the GVRH and Band Gap datasets, respectively. However, DSSL+VE does not exhibit performance improvement over DSSL due to the previously analyzed reason: valence electrons lack a strong correlation with formation energy, making it challenging to guide the prediction of formation energy. Then for the HOIP dataset, we find that both Mask SSL and contrastive SSL lead to worse performance compared to the Original model with an MAE of 0.237 eV compared to 0.214 eV. The same is true for the Lanthanides dataset. This performance degradation by SSL is, as we analyzed before, due to the fact that the pretraining data from the MP datasets are significantly different from these two downstream task datasets. Here, the pretraining on the MP dataset becomes a hindrance to the finetuning stage, so that the original DeeperGATGNN has the best results. 

Additionally, we applied the physics-guided SSL strategy over another SOTA network, ALIGNN, with the results shown in Table \ref{tab:comp_alignn}. We initially trained the original ALIGNN on the GVRH and KVRH datasets using five-fold cross-validation, resulting in average MAEs of 0.0720 GPa and 0.0574 GPa, respectively, as shown in the first row of Table \ref{tab:comp_alignn}. Then we pretrained an ALIGNN model with the same hyperparameters as above, incorporating the prediction of atomic stiffness as the pretext task on the MP dataset. Following pretraining, we finetuned this pretrained model on the GVRH and KVRH datasets using five-fold cross-validation. This variant of ALIGNN with the pretraining process, involving the prediction of atomic stiffness, is denoted as "ALIGNN+AS". In comparison with the original ALIGNN, ALIGNN+AS demonstrated average MAEs of 0.0683 GPa and 0.0557 GPa, respectively, indicating improvements of 5.56\% and 2.96\% on the GVRH and KVRH datasets, respectively.

\begin{table}[th]
\caption{Performance comparison of different SSL strategies. From top to bottom, each row of the table represents the MAE performance of the original DeeperGATGNN, the network pretrained with only masks strategy (MASK), the network pretrained with only contrastive learning strategy (CL), and our DSSL network.}
\label{tab:comp_ssl}
\begin{tabular}{c|cccccc}
\toprule[1.5pt]
             & GVRH                   & Band Gap               & Formation Energy       & Fermi Energy           & HOIP                   & Lanthanides            \\ \hline
Original & 0.095 ± 0.003          & 0.300 ± 0.011          & 0.023 ± 0.001          & 0.374 ± 0.014          & \textbf{0.214 ± 0.017} & \textbf{0.105 ± 0.008} \\
Mask         & 0.085 ± 0.003          & 0.297 ± 0.008 & 0.023 ± 0.001          & 0.367 ± 0.010            & 0.237 ± 0.026          & 0.138 ± 0.014          \\
CL           & 0.154 ± 0.046          & 0.301 ± 0.005          & 0.023 ± 0.001          & 0.385 ± 0.012          & 0.237 ± 0.018          & 0.134 ± 0.011          \\
DSSL         & 0.084 ± 0.003 & 0.289 ± 0.005         & \textbf{0.021 ± 0.001} & \textbf{0.351 ± 0.005} & 0.232 ± 0.015          & 0.154 ± 0.015          \\
DSSL+Micro         & \textbf{0.083 ± 0.003} & \textbf{0.283 ± 0.005}          & 0.022 ± 0.001 & - & -  & -          \\
\bottomrule[1.5pt]  
\end{tabular}
\end{table}

\begin{table}[th]
\centering
\caption{Performance comparison of the original ALIGNN and the ALIGNN that uses Atomic Stiffness as the pretraining task (ALIGNN+AS) on GVRH and KVRH datasets.}
\label{tab:comp_alignn}
\begin{tabular}{c|cc}
\toprule[1.5pt]
          & GVRH           & KVRH                    \\ \hline
ALIGNN    & 0.0720 ± 0.001  & 0.0574 ± 0.001          \\
ALIGNN+AS & 0.0683 ± 0.002 & 0.0557 ± 0.001 \\
\bottomrule[1.5pt]  
\end{tabular}
\end{table}

\section{Conclusion}
We introduce a physics-guided GNN-based dual SSL framework, DSSL for structure based materials property prediction. The framework contains two stages. In the pretraining stage, the model is trained on the MP dataset using both mask-based and contrastive learning SSL strategies to capture both local and global information for the input crystal structures. Additionally, we propose using macroproperty related microproperty prediction as additional pretext tasks in the pretraining stage to achieve physics-guided neural network training. From the experiments, we observed that our DSSL can achieve better performance than its backbone model, DeeperGATGNN, and the majority of SOTA graph neural network models in predicting various material properties. However, we find that for downstream tasks with different data distributions from the pertraining dataset, it is hard for DSSL to improve the performance of the baseline model. 
We further designed the ablation study to examine the impact of different SSL strategies on the performance of the baseline GNN model and found that our DSSL with physics-guided pretext tasks can achieve the best performance as shown by the band gap and elasticity prediction problems. However, similar challenges arise when dealing with datasets with different distributions from the pretraining dataset. Given this problem, it may have to adjust or expand the pertaining dataset according to the specific downstream task of property prediction. 

Our work may be further improved by exploiting the newly discovered 2.2 million stable structures by the DeepMind team \cite{merchant2023scaling}, which have significantly expanded the stable materials known to humanity via elemental substitution and crystal structure prediction. 
Moreover, we want to emphasize that our DSSL framework can naturally include additional pretext tasks such as the force and formation energy prediction into the pretraining stage to further improve the model performance since these two physical quantities are fundamental to many macroproperties and have played a key role in a related SSL study \cite{shoghi2023molecules}.  We also apply our physics-guided SSL strategy to ALLIGAN, another SOTA network model and observed improved performance. Together, we demonstrate that self-supervised learning provides an effective approach to injecting physical concepts into neural network models to ensure better prediction performance.

\section{Data and Code Availability}
The source code and datasets can be freely accessed at \url{https://github.com/usccolumbia/DSSL}.

\section{Contribution}
Conceptualization, J.H.; methodology, N.F.,J.H. ; software, N.F.; resources, J.H.; writing--original draft preparation, N.F.,J.H.; writing--review and editing, J.H., L.W.; visualization, N.F.; supervision, J.H.;  funding acquisition, J.H.

\section*{Acknowledgement}
The research reported in this work was supported in part by National Science Foundation under the grant and 10011239,10013417, and 10013216. The views, perspectives, and content do not necessarily represent the official views of the NSF.

\bibliographystyle{unsrt}  
\bibliography{references}

\begin{thebibliography}{10}

\bibitem{le2012quantitative}
Tu~Le, V~Chandana Epa, Frank~R Burden, and David~A Winkler.
\newblock Quantitative structure--property relationship modeling of diverse
  materials properties.
\newblock {\em Chemical reviews}, 112(5):2889--2919, 2012.

\bibitem{hu2022piezoelectric}
Jeffrey Hu and Yuqi Song.
\newblock Piezoelectric modulus prediction using machine learning and graph
  neural networks.
\newblock {\em Chemical Physics Letters}, 791:139359, 2022.

\bibitem{varivoda2023materials}
Daniel Varivoda, Rongzhi Dong, Sadman~Sadeed Omee, and Jianjun Hu.
\newblock Materials property prediction with uncertainty quantification: A
  benchmark study.
\newblock {\em Applied Physics Reviews}, 10(2), 2023.

\bibitem{fung2021benchmarking}
Victor Fung, Jiaxin Zhang, Eric Juarez, and Bobby~G Sumpter.
\newblock Benchmarking graph neural networks for materials chemistry.
\newblock {\em npj Computational Materials}, 7(1):1--8, 2021.

\bibitem{xie2018crystal}
Tian Xie and Jeffrey~C Grossman.
\newblock Crystal graph convolutional neural networks for an accurate and
  interpretable prediction of material properties.
\newblock {\em Physical review letters}, 120(14):145301, 2018.

\bibitem{chen2019graph}
Chi Chen, Weike Ye, Yunxing Zuo, Chen Zheng, and Shyue~Ping Ong.
\newblock Graph networks as a universal machine learning framework for
  molecules and crystals.
\newblock {\em Chemistry of Materials}, 31(9):3564--3572, 2019.

\bibitem{louis2020graph}
Steph-Yves Louis, Yong Zhao, Alireza Nasiri, Xiran Wang, Yuqi Song, Fei Liu,
  and Jianjun Hu.
\newblock Graph convolutional neural networks with global attention for
  improved materials property prediction.
\newblock {\em Physical Chemistry Chemical Physics}, 22(32):18141--18148, 2020.

\bibitem{choudhary2021atomistic}
Kamal Choudhary and Brian DeCost.
\newblock Atomistic line graph neural network for improved materials property
  predictions.
\newblock {\em npj Computational Materials}, 7(1):185, 2021.

\bibitem{omee2022scalable}
Sadman~Sadeed Omee, Steph-Yves Louis, Nihang Fu, Lai Wei, Sourin Dey, Rongzhi
  Dong, Qinyang Li, and Jianjun Hu.
\newblock Scalable deeper graph neural networks for high-performance materials
  property prediction.
\newblock {\em Patterns}, 3(5):100491, 2022.

\bibitem{schrier2023pursuit}
Joshua Schrier, Alexander~J Norquist, Tonio Buonassisi, and Jakoah Brgoch.
\newblock In pursuit of the exceptional: Research directions for machine
  learning in chemical and materials science.
\newblock {\em Journal of the American Chemical Society}, 145(40):21699--21716,
  2023.

\bibitem{ericsson2022self}
Linus Ericsson, Henry Gouk, Chen~Change Loy, and Timothy~M Hospedales.
\newblock Self-supervised representation learning: Introduction, advances, and
  challenges.
\newblock {\em IEEE Signal Processing Magazine}, 39(3):42--62, 2022.

\bibitem{liu2021self}
Xiao Liu, Fanjin Zhang, Zhenyu Hou, Li~Mian, Zhaoyu Wang, Jing Zhang, and Jie
  Tang.
\newblock Self-supervised learning: Generative or contrastive.
\newblock {\em IEEE transactions on knowledge and data engineering},
  35(1):857--876, 2021.

\bibitem{cao2023moformer}
Zhonglin Cao, Rishikesh Magar, Yuyang Wang, and Amir Barati~Farimani.
\newblock Moformer: self-supervised transformer model for metal--organic
  framework property prediction.
\newblock {\em Journal of the American Chemical Society}, 145(5):2958--2967,
  2023.

\bibitem{kolluru2022transfer}
Adeesh Kolluru, Nima Shoghi, Muhammed Shuaibi, Siddharth Goyal, Abhishek Das,
  C~Lawrence Zitnick, and Zachary Ulissi.
\newblock Transfer learning using attentions across atomic systems with graph
  neural networks (taag).
\newblock {\em The Journal of Chemical Physics}, 156(18), 2022.

\bibitem{doersch2017multi}
Carl Doersch and Andrew Zisserman.
\newblock Multi-task self-supervised visual learning.
\newblock In {\em Proceedings of the IEEE international conference on computer
  vision}, pages 2051--2060, 2017.

\bibitem{wu2021self}
Lirong Wu, Haitao Lin, Cheng Tan, Zhangyang Gao, and Stan~Z Li.
\newblock Self-supervised learning on graphs: Contrastive, generative, or
  predictive.
\newblock {\em IEEE Transactions on Knowledge and Data Engineering}, 2021.

\bibitem{magar2022crystal}
Rishikesh Magar, Yuyang Wang, and Amir Barati~Farimani.
\newblock Crystal twins: self-supervised learning for crystalline material
  property prediction.
\newblock {\em npj Computational Materials}, 8(1):231, 2022.

\bibitem{zbontar2021barlow}
Jure Zbontar, Li~Jing, Ishan Misra, Yann LeCun, and St{\'e}phane Deny.
\newblock Barlow twins: Self-supervised learning via redundancy reduction.
\newblock In {\em International Conference on Machine Learning}, pages
  12310--12320. PMLR, 2021.

\bibitem{chen2021exploring}
Xinlei Chen and Kaiming He.
\newblock Exploring simple siamese representation learning.
\newblock In {\em Proceedings of the IEEE/CVF conference on computer vision and
  pattern recognition}, pages 15750--15758, 2021.

\bibitem{das2023crysgnn}
Kishalay Das, Bidisha Samanta, Pawan Goyal, Seung-Cheol Lee, Satadeep
  Bhattacharjee, and Niloy Ganguly.
\newblock Crysgnn: Distilling pre-trained knowledge to enhance property
  prediction for crystalline materials.
\newblock {\em arXiv preprint arXiv:2301.05852}, 2023.

\bibitem{suzuki2022self}
Yuta Suzuki, Tatsunori Taniai, Kotaro Saito, Yoshitaka Ushiku, and Kanta Ono.
\newblock Self-supervised learning of materials concepts from crystal
  structures via deep neural networks.
\newblock {\em Machine Learning: Science and Technology}, 3(4):045034, 2022.

\bibitem{shoghi2023molecules}
Nima Shoghi, Adeesh Kolluru, John~R Kitchin, Zachary~W Ulissi, C~Lawrence
  Zitnick, and Brandon~M Wood.
\newblock From molecules to materials: Pre-training large generalizable models
  for atomic property prediction.
\newblock {\em arXiv preprint arXiv:2310.16802}, 2023.

\bibitem{OC20}
Lowik Chanussot, Abhishek Das, Siddharth Goyal, Thibaut Lavril, Muhammed
  Shuaibi, Morgane Riviere, Kevin Tran, Javier Heras-Domingo, Caleb Ho, Weihua
  Hu, et~al.
\newblock Open catalyst 2020 (oc20) dataset and community challenges.
\newblock {\em Acs Catalysis}, 11(10):6059--6072, 2021.

\bibitem{OC22}
Richard Tran, Janice Lan, Muhammed Shuaibi, Brandon~M Wood, Siddharth Goyal,
  Abhishek Das, Javier Heras-Domingo, Adeesh Kolluru, Ammar Rizvi, Nima Shoghi,
  et~al.
\newblock The open catalyst 2022 (oc22) dataset and challenges for oxide
  electrocatalysts.
\newblock {\em ACS Catalysis}, 13(5):3066--3084, 2023.

\bibitem{ANI-1}
Justin~S Smith, Olexandr Isayev, and Adrian~E Roitberg.
\newblock Ani-1, a data set of 20 million calculated off-equilibrium
  conformations for organic molecules.
\newblock {\em Scientific data}, 4(1):1--8, 2017.

\bibitem{Transition1x}
Mathias Schreiner, Arghya Bhowmik, Tejs Vegge, Jonas Busk, and Ole Winther.
\newblock Transition1x-a dataset for building generalizable reactive machine
  learning potentials.
\newblock {\em Scientific Data}, 9(1):779, 2022.

\bibitem{jin2023atomic}
Ruihua Jin, Xiaoang Yuan, and Enlai Gao.
\newblock Atomic stiffness for bulk modulus prediction and high-throughput
  screening of ultraincompressible crystals.
\newblock {\em Nature Communications}, 14(1):4258, 2023.

\bibitem{Sirdeshmukh2006MicroAM}
Dinker~B. Sirdeshmukh, Lalitha Sirdeshmukh, and K.~G. Subhadra.
\newblock Micro- and macro-properties of solids : thermal, mechanical and
  dielectric properties.
\newblock 2006.

\bibitem{10.1063/1.4812323}
Anubhav Jain, Shyue~Ping Ong, Geoffroy Hautier, Wei Chen, William~Davidson
  Richards, Stephen Dacek, Shreyas Cholia, Dan Gunter, David Skinner, Gerbrand
  Ceder, and Kristin~A. Persson.
\newblock {Commentary: The Materials Project: A materials genome approach to
  accelerating materials innovation}.
\newblock {\em APL Materials}, 1(1), 07 2013.
\newblock 011002.

\bibitem{dunn2020benchmarking}
Alexander Dunn, Qi~Wang, Alex Ganose, Daniel Dopp, and Anubhav Jain.
\newblock Benchmarking materials property prediction methods: the matbench test
  set and automatminer reference algorithm.
\newblock {\em npj Computational Materials}, 6(1):138, 2020.

\bibitem{le2020contrastive}
Phuc~H Le-Khac, Graham Healy, and Alan~F Smeaton.
\newblock Contrastive representation learning: A framework and review.
\newblock {\em Ieee Access}, 8:193907--193934, 2020.

\bibitem{jain2013commentary}
Anubhav Jain, Shyue~Ping Ong, Geoffroy Hautier, Wei Chen, William~Davidson
  Richards, Stephen Dacek, Shreyas Cholia, Dan Gunter, David Skinner, Gerbrand
  Ceder, et~al.
\newblock Commentary: The materials project: A materials genome approach to
  accelerating materials innovation.
\newblock {\em APL materials}, 1(1):011002, 2013.

\bibitem{deJong2015}
Maarten de~Jong, Wei Chen, Thomas Angsten, Anubhav Jain, Randy Notestine,
  Anthony Gamst, Marcel Sluiter, Chaitanya Krishna~Ande, Sybrand van~der Zwaag,
  Jose~J. Plata, Cormac Toher, Stefano Curtarolo, Gerbrand Ceder, Kristin~A.
  Persson, and Mark Asta.
\newblock Charting the complete elastic properties of inorganic crystalline
  compounds.
\newblock {\em Scientific Data}, 2:150009, Mar 2015.
\newblock Data Descriptor.

\bibitem{materialsproject}
Anubhav Jain, Shyue~Ping Ong, Geoffroy Hautier, Wei Chen, William~Davidson
  Richards, Stephen Dacek, Shreyas Cholia, Dan Gunter, David Skinner, Gerbrand
  Ceder, et~al.
\newblock Commentary: The materials project: A materials genome approach to
  accelerating materials innovation.
\newblock {\em APL Materials}, 1(1):011002, 2013.

\bibitem{karamad2020orbital}
Mohammadreza Karamad, Rishikesh Magar, Yuting Shi, Samira Siahrostami, Ian~D
  Gates, and Amir~Barati Farimani.
\newblock Orbital graph convolutional neural network for material property
  prediction.
\newblock {\em Physical Review Materials}, 4(9):093801, 2020.

\bibitem{kim2017hybrid}
Chiho Kim, Tran~Doan Huan, Sridevi Krishnan, and Rampi Ramprasad.
\newblock A hybrid organic-inorganic perovskite dataset.
\newblock {\em Scientific data}, 4(1):1--11, 2017.

\bibitem{pham2017machine}
Tien~Lam Pham, Hiori Kino, Kiyoyuki Terakura, Takashi Miyake, Koji Tsuda,
  Ichigaku Takigawa, and Hieu~Chi Dam.
\newblock Machine learning reveals orbital interaction in materials.
\newblock {\em Science and technology of advanced materials}, 18(1):756, 2017.

\bibitem{petretto2018high}
Guido Petretto, Shyam Dwaraknath, Henrique PC~Miranda, Donald Winston, Matteo
  Giantomassi, Michiel~J Van~Setten, Xavier Gonze, Kristin~A Persson, Geoffroy
  Hautier, and Gian-Marco Rignanese.
\newblock High-throughput density-functional perturbation theory phonons for
  inorganic materials.
\newblock {\em Scientific data}, 5(1):1--12, 2018.

\bibitem{petousis2017high}
Ioannis Petousis, David Mrdjenovich, Eric Ballouz, Miao Liu, Donald Winston,
  Wei Chen, Tanja Graf, Thomas~D Schladt, Kristin~A Persson, and Fritz~B Prinz.
\newblock High-throughput screening of inorganic compounds for the discovery of
  novel dielectric and optical materials.
\newblock {\em Scientific data}, 4(1):1--12, 2017.

\bibitem{XIAO2024112619}
Jianping Xiao, Li~Yang, and Shuqun Wang.
\newblock Graph isomorphism network for materials property prediction along
  with explainability analysis.
\newblock {\em Computational Materials Science}, 233:112619, 2024.

\bibitem{merchant2023scaling}
Amil Merchant, Simon Batzner, Samuel~S. Schoenholz, Muratahan Aykol, Gowoon
  Cheon, and Ekin~Dogus Cubuk.
\newblock Scaling deep learning for materials discovery.
\newblock {\em Nature}, 2023.

\end{thebibliography}

\end{document}